\documentclass[%
 reprint,
 amsmath,amssymb,
 aps,
]{revtex4-1}

\usepackage{graphicx}
\usepackage{dcolumn}
\usepackage{bm}
\usepackage[version=3]{mhchem} 
\usepackage[T1]{fontenc}       
\usepackage{hyperref}
\usepackage{amssymb}
\usepackage{graphicx}
\usepackage{epstopdf}
\usepackage{cleveref}
\usepackage{booktabs, array}
\usepackage{multirow}
\usepackage{mathtools, bm, nicefrac, dsfont}
\usepackage{subfigure}
\usepackage{siunitx}
\usepackage{afterpage}
\usepackage{lipsum}
\usepackage{tikz}
	\usetikzlibrary{automata,positioning,shapes,decorations,matrix}
\usepackage{chemfig}
\usepackage{siunitx}
	\DeclareSIUnit{\tss}{\text{timesteps}}
\usepackage{bigstrut}
\usepackage{enumitem}
\usepackage{nicefrac}

\definecolor{MyBlue}{RGB}{55,126,184}
\definecolor{MyRed}{RGB}{228,26,28}
\definecolor{MyGreen}{RGB}{77,175,74}

\newcommand{\revision}[1]{\textcolor{black}{#1}}

\newcommand{\llzo}[0]{$\text{Li}_7\text{La}_3\text{Zr}_2\text{O}_{12}$}
\newcommand{\li}[0]{\ce{Li^+}}
\newcommand{\emim}[0]{\ce{[Emim]^+}}
\newcommand{\tfo}[0]{\ce{[TFO]^-}}
\newcommand{\phosphate}[0]{\ce{Li_3PO_4}}

\begin{document}

\title{Spectral denoising for accelerated analysis of correlated ionic transport}

\author{Nicola Molinari}
    \email{nmolinari@seas.harvard.edu}
\author{Yu Xie}
\author{Ian Leifer}
\affiliation{%
    John A. Paulson School of Engineering and Applied Sciences, \\
    Harvard University, Cambridge, MA 02138, USA.
}%
\author{Aris Marcolongo}
\affiliation{Universit\"{a}t Bern, Bern CH-3012, Switzerland.}
\author{Mordechai Kornbluth}
\affiliation{%
    Robert Bosch LLC, Research and Technology Center, \\
    Cambridge, Massachusetts 02142, USA.}
\author{Boris Kozinsky}
\email{bkoz@seas.harvard.edu}
\affiliation{%
    John A. Paulson School of Engineering and Applied Sciences, \\
    Harvard University, Cambridge, MA 02138, USA.
}%
\affiliation{%
    Robert Bosch LLC, Research and Technology Center, \\
    Cambridge, Massachusetts 02142, USA.}

\date{\today}

\begin{abstract}
\noindent Computation of correlated ionic transport properties from molecular dynamics in the Green-Kubo formalism is expensive as one cannot rely on the affordable mean square displacement approach. We use spectral decomposition of the short-time ionic displacement covariance to learn a set of diffusion eigenmodes that encode the correlation structure and form a basis for analyzing the ionic trajectories. This allows to systematically reduce the uncertainty and accelerate computations of ionic conductivity in systems with a steady-state correlation structure. We provide mathematical and numerical proofs of the method's robustness, and demonstrate it on realistic electrolyte materials.
\end{abstract}


\maketitle


\section{\label{sec:level1}Introduction}
\indent Understanding the transport of ionic species is of central importance in a variety of fields ranging from physics and biophysics\cite{goodenough2004electronic, fulinski1997active} to chemistry in general\cite{bachman2016inorganic}.
Of particular relevance for energy storage solutions, the design of next-generation metal-ion batteries depends on the development of fast ion-conducting and stable electrolyte materials\cite{schmuch2018performance, xu2014electrolytes, balduccicritical}.

\indent As the total ionic conductivity is proportional to the number of charge carriers, high concentrations of ions are typically targeted.
Consequently, due to both high concentrations and long-range Coulomb interactions, the correlation between ions becomes non-negligible\cite{qiao2018supramolecular, molinari2019transport}.
{%
The physics of ion correlations and effects on transport properties are complex as they can be either beneficial\cite{he2017origin, murch1982haven} or detrimental\cite{molinari2019general, gouverneur2018negative} depending on the composition and concentrations.
To capture these effects one can either apply an external electric field, or collect the statistics of the total ionic flux fluctuation at equilibrium\cite{wheeler2004molecular, wheeler2004moleculara}.
The former (non-equilibrium) method is often problematic, especially for first-principles dynamics simulations, due to the difficulty of including a finite electric field in periodic systems, ensuring its linear-response effect, and controlling the thermodynamic ensemble of the driven system. \cite{maginn1993transport, wheeler2004moleculara}.
Consequently, the latter (equilibrium) method, based on the Green-Kubo formalism\cite{kubo1957statistical}, is often preferred. Even though the method is exact in principle, the challenge is that it is based on fluctuations of a single total flux value for the entire system, and the conductivity estimate has a variance that increases with the system size.
As a result, to reach sufficient convergence of the statistics, long simulations are needed, especially for large systems where total flux fluctuations are small.
}

{These limitations often lead to the erroneous adoption of the Nernst-Einstein dilute-solution approximation.}
In this case the variance of the estimate is lower and independent of the system size, leading to fast although often incorrect (biased) estimates.
If the short-range correlations are strong, time-independent, and known a priori, such as bonds between atoms in a molecular liquid, with negligible inter-molecular correlations, displacements of the molecules rather than atoms can be used in the Nernst-Einstein formulation.
{%
In the case of electrolytes, for example, the cluster Nernst-Einstein method treats ionic clusters as uncorrelated charge-moving entities \cite{france2019correlations}.
Equivalent to the Nernst-Einstein approach for uncorrelated molecular diffusion, this approach reduces the $\kappa$ estimate variance.
However, it relies on the prior knowledge of the clusters' composition and the assumption of their immutability, which is not satisfied by a wide range of liquid, polymer and single-ion solid-state electrolytes.
In the absence of perfect knowledge of clusters, or dynamic nature of the interatomic correlations, the expensive total flux approach is required.
}

\indent In this work, we introduce a data-driven approach to learn the diffusive modes from the full interatomic displacement covariance matrix via its spectral decomposition, without prior assumptions about the range and structure of the correlation, other than it is time-independent. We then use the learned eigenvectors to greatly reduce the variance and accelerate the estimation of the ionic conductivity for correlated systems, with rigorous provable bounds on the bias and variance.

\section{\label{sec:level1}Theory \& Method}
The conductivity $\kappa$ can be written in an Einstein form as:
\begin{align}
    \kappa = \dfrac{F^2}{6\mathcal{N}^2_{\rm A} V k_B \mathcal{T}} \lim_{\tau \rightarrow \infty}  \dfrac{\partial}{\partial\tau} \sum_{ij} q_iq_j \langle\mathcal{C}_{ij}(\tau)\rangle. \label{eqn:einstein}
\end{align}
Here $F$ is the Faraday's constant, $\mathcal{N}_{\rm A}$ the Avogadro's number, $V$ the volume, $k_B$ the Boltzmann's constant, $\mathcal{T}$ the temperature, $\langle\cdots\rangle$ indicates an element-by-element average (see below), and $q_i$ is the charge and $\mathbf{r}_i\left(t\right)$ the position vector of particle $i$ at time $t$.
The position displacement covariance matrix is $\mathcal{C}_{ij}(\tau)=\left[ \mathbf{r}_i\left(\tau + t\right) - \mathbf{r}_i\left(t\right) \right]\cdot \left[ \mathbf{r}_j\left(\tau+t\right) - \mathbf{r}_j\left(t\right) \right]$ and indices $i,j$ run over all particles.

\indent The diagonal $i = j$ represents each particle's squared displacements, and the inter-particle displacement correlation is encoded in the off-diagonal $i \neq j$ elements.
If the motions of particles $i$ and $j$ are uncorrelated, the averaged, off-diagonal element $\langle \mathcal{C}_{i\neq j}(\tau)\rangle \rightarrow 0$ for values of $\tau$ greater than the typical collision time scale.
Consequently, the relevant information about the diffusion of an uncorrelated system lies exclusively on the diagonal of $\langle \mathcal{C}_{ij}(\tau)\rangle$.
Replacing the $\langle\mathcal{C}_{i\neq j}\left(\tau\right)\rangle$ terms with zeros is equivalent to the widely-adopted Nernst-Einstein formulation\cite{mott1940electronic}, which is only correct in the infinitely dilute uncorrelated limit.
In this case each diagonal value contributes independently to the variance of the conductivity estimate, which is reduced by a factor of $N$ (number of particles) compared to the variance of the total flux estimate of the conductivity\cite{muller1995computer}.
Finally, the $\langle \mathcal{C}_{ij}(\tau) \rangle$ average is obtained by performing multiple instances of the same experiment, and by implementing time-window averaging\cite{allen2017computer}.
{The issue with the latter averaging is that the statistics becomes poorer for longer time-windows, i.e., if $\Delta t_1 < \Delta t_2$ then $\langle \mathcal{C}_{ij}(\Delta t_1) \rangle$ is averaged over more matrices, thus better converged, than $\langle \mathcal{C}_{ij}(\Delta t_2) \rangle$.}
Thus, determination of transport properties becomes noisier, and long MD simulations are needed to obtain the total flux conductivity estimate within a reasonable uncertainty. \\

\indent Our approach is based on the eigenvectors of $\langle \mathcal{C}_{ij}(\tau)\rangle$ averaged using a short time interval $\tau = \tau_1$  greater than the minimum time needed for the system to reach the diffusive regime. $\langle \mathcal{C}_{ij}(\tau_1)\rangle$ contains the most well-converged information about the correlation of the system because any time window $\tau_n > \tau_1$ has fewer position-position correlation matrices to average over.
We then use the learned spectral information to reduce the statistical variance of the estimate of $\kappa$ for any time window $\tau_n > \tau_1$.
{The only assumption of our approach is that the correlation structure of the system under investigation does not change over time.
{If a system is in steady-state with respect to the diffusion timescale, the correlation structure is determined by the chemistry at play as well as the distribution of diffusion mechanisms available at a given temperature.}
}

\indent \revision{Before describing the spectral denoising approach, we clarify that, for our purposes, tracer and self-diffusion are used interchangeably.
However, we highlight that the two quantities might differ when same-particle successive diffusion jumps are correlated, such as in vacancy-mediated diffusion (not discussed here)\cite{DiffusionInSolids, howard1964matter}.}
Additionally, we introduce the following terminology used throughout this work.
\begin{itemize}[leftmargin=0 em]
    \item[] \textbf{Total flux}, or full summation (FS) is the exact Green-Kubo conductivity based on the summation over \textit{all} $i,j$ elements of $\langle\mathcal{C}_{ij}(\tau)\rangle$ as in \autoref{eqn:einstein}.
    \item[] {\textbf{Nernst-Einstein} self-diffusion (or MSD) approach based on the trace of $\langle\mathcal{C}_{ij}(\tau)\rangle$, assumes the cross-terms $i \neq j$ are zero. Nernst-Einstein conductivity is only correct for infinitely-dilute systems.}
    \item[] \textbf{Spectral denoising} (SD) refers to the method developed in this work, where the denoised $\langle\mathcal{C}^*_{ij}(\tau)\rangle$ is used to compute conductivity, as described below.
\end{itemize}
{To quantify the degree of correlation in a given system, the correlation factor $f_c=\nicefrac{\kappa_{\rm FS}}{\kappa_{\rm MSD}}= \nicefrac{\sum_{ij}\mathcal{C}_{ij}}{\sum_i \mathcal{C}_{ii}}$ is introduced as the ratio of the FS to MSD estimate of conductivity (it is the inverse of the Haven ratio, and it is different from the same-particle correlation factor that quantifies the departure from a random walk).}

\indent Now we outline the spectral denoising approach to reduce the noise in the calculation of correlated ionic conductivity.
(1) Diagonalize $\langle\mathcal{C}_{ij}(\tau_1)\rangle$, and obtain its eigenbasis $\mathcal{A}$.
As $\langle\mathcal{C}_{ij}(\tau_1)\rangle$ is positive-definite and symmetric, it is always possible to find a complete set of real, orthonormal eigenvectors, and its eigenvalues are positive.
(2) Rotate using $\mathcal{A}$ all $\langle\mathcal{C}_{ij}(\tau_n)\rangle$, where $\tau_n > \tau_1$, obtaining $\Gamma(\tau_n) = \mathcal{A}^{T}\langle\mathcal{C}_{ij}(\tau_n)\rangle\mathcal{A}$.
As we assume the system is in equilibrium and the correlation profile is stationary, we expect $\Gamma(\tau_n)$ to be nearly diagonal, with noise in the off-diagonal terms only arising from finite sampling of the time-window averages.
This noise is confirmed numerically to have zero expectation value, Section 4 of the Supplemental Material\cite{molinari2020supplemental}.
(3) Set to zero the off-diagonal terms of $\Gamma(\tau_n)$ obtaining $\Gamma^*(\tau_n)$, in the same spirit as in the Nernst-Einstein. The key difference is that we do this in the basis of natural diffusion eigenmodes identified from the covariance matrix of the system itself.
(4) Rotate back $\Gamma^*(\tau_n)$ using the eigenbasis $\mathcal{A}$ to obtain the denoised covariance matrix $\langle\mathcal{C}^*_{ij}(\tau_n)\rangle$ to then use in~\autoref{eqn:einstein}.
In addition to helping reduce the conductivity variance, the eigenbasis $\mathcal{A}$ also provides a microscopic understanding of the fundamental modes of diffusion in correlated scenarios, as discussed below.
The details regarding the choice of the optimal $\tau_1$, as well as the mathematical robustness of this approach are thoroughly discussed in Section 3 of the Supplemental Material\cite{molinari2020supplemental}.
{We rigorously prove that the SD method outperforms the FS method if the eigenbasis $\mathcal{A}$ is ideal, i.e., calculated from the true covariance matrix obtained with infinite statistics of the diffusion process.}
In this limit the SD approach yields unbiased results, while the estimate variance is always strictly lower than that of FS:
\begin{align}
    \operatorname{Var}\left(\sum_{i j} \mathcal{C}^*_{i j}\right)
    &=2 \sum_{i}\left(\lambda_{i} w_{i}^{2}\right)^{2} 
    \nonumber \\
    &\leq 2 \left(\sum_{i}\lambda_{i} w_{i}^{2}\right)^{2} 
    = \operatorname{Var}\left(\sum_{i j} \mathcal{C}_{i j}\right). \nonumber
\end{align}
Where $\lambda_i \geq 0$ are the diagonal elements of the $\Gamma$ matrix, and $w_i$ is the sum of all elements of the $i$th eigenvector, i.e., column of $\mathcal{A}$.
In practice, $\left\langle\mathcal{C}_{i j}\left(\tau_{1}\right)\right\rangle$ is a finite-sampling approximation to the ideal covariance matrix.
In this case, we can also prove that the SD approach outperforms the FS - full details of the proofs and derivations are given in Section 3 of the Supplemental Material\cite{molinari2020supplemental}.

\section{\label{sec:level1}Validation \& Applications}
\subsubsection{\label{sec:level3}Multivariate Gaussian Random Walk}
In order establish the methodology in a controlled model setting, we sample a multivariate Gaussian distribution to obtain correlated random displacement vectors that mimic Brownian diffusion with exactly known correlation.
{We consider a model with the covariance matrix with $\mathcal{C}_{i=j}=\alpha$ and $\mathcal{C}_{i\neq j}=\beta$ that represents a homogeneous single-component correlated system.
The correlation factor is then defined as $f_c = {\sum_{ij}\mathcal{C}_{ij} \over \sum_i \mathcal{C}_{ii}} = {N\alpha+N(N-1)\beta \over N\alpha}.$
A system with $f_c=1$ possesses no correlation (the sum of the off-diagonal covariance elements equals \num{0}, in this case $\beta \equiv 0$), while $f_c>1$ and $0<f_c<1$ corresponds to systems with negative and positive inter-particle correlation, respectively.
Thus, we expect the SD approach to provide the greatest improvement over the FS for $f_c=1$, while reducing to FS for $f_c \ll 1$ and $f_c \gg 1$.
More details are provided in Section 1 of the Supplemental Material\cite{molinari2020supplemental}.}
A time series of $n$ Brownian steps is then constructed by adding $n$ sampled vectors, and our goal is to estimate the slope $D = \frac{\partial}{\partial\tau}\sum_{ij}\langle \mathcal{C}_{ij}\left(\tau\right)\rangle$ from the sampled displacement trajectories.
By modifying the covariance matrix, we test the applicability and robustness of our approach on a wide range of types and strengths of correlations.
More details on the Gaussian sampling are discussed in Section 2 of the Supplemental Material\cite{molinari2020supplemental}.

\indent For our testing, we set the number of Brownian steps to \num{1000}, and a minimum of \num{10} averaged position-position correlation matrices $\langle\mathcal{C}_{ij}(\tau_n)\rangle$ to perform linear regression to obtain the slope $D$.
The number $\mathit{N}$ of simulated Brownian walkers, representing diffusing particles, ranges from \numrange{3}{500}.
We study $f_c$ in the range of \numrange{0.25}{2.75}.
%
%
For every $(f_c, \mathit{N})$ combination, we perform \num{100} independent simulations of the Brownian walkers.

\begin{figure}[h!tb]
    \centering
    \includegraphics[scale=1.0]{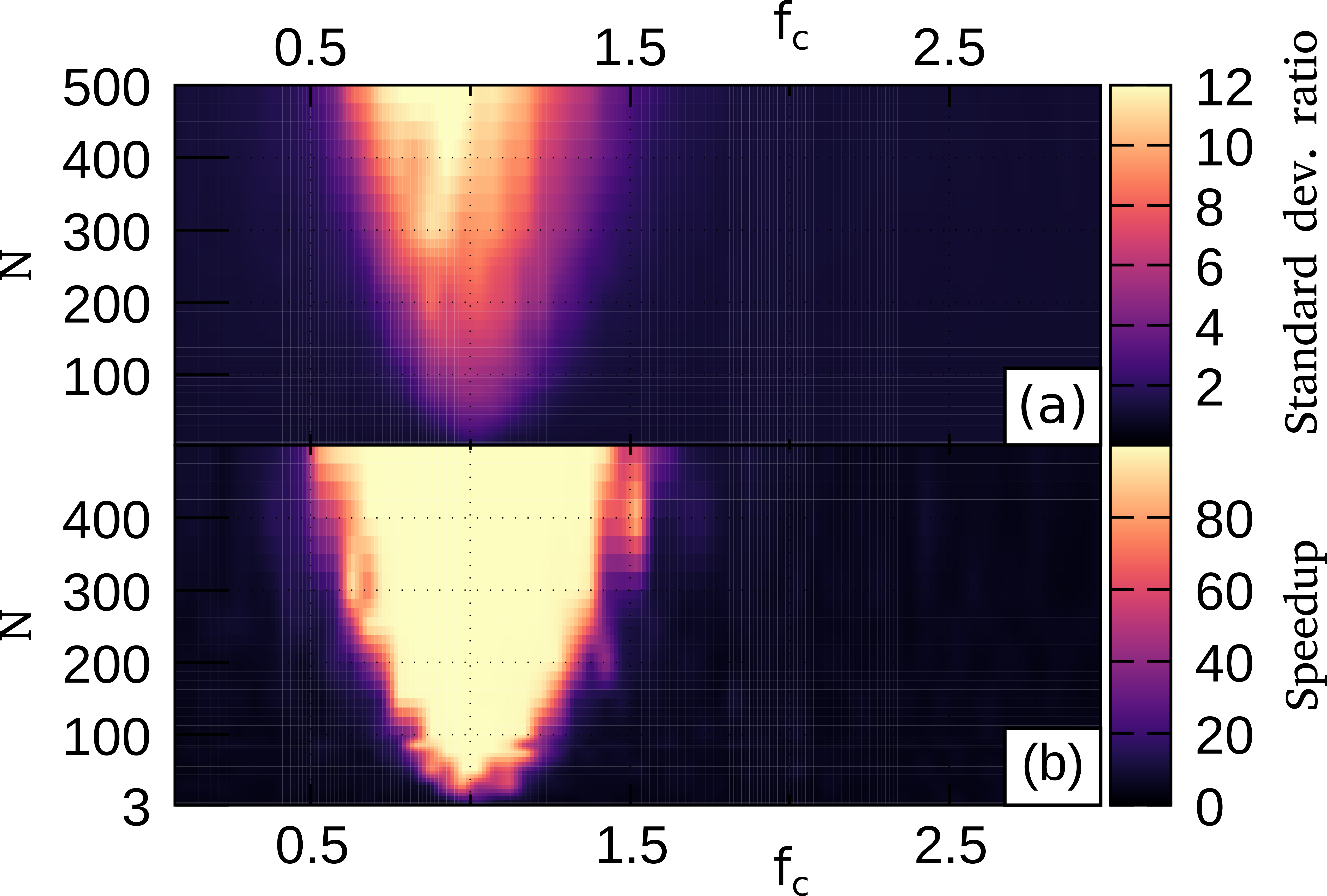}
    \caption{Improvement of the SD estimates of the slope $D$ of the multivariate Gaussian walk model: (a), the standard deviation reduction and, (b), the computational speedup of the SD method, relative to the FS method.}
    \label{fig:fig1}
\end{figure}

\indent \autoref{fig:fig1} shows two ways the SD approximation can be used to improve the determination of correlated transport properties with respect to the exact theory.
Panel (a) presents the reduction in the standard deviation of the $D$ estimate, calculated as the ratio between the standard error of the FS and that of the SD approach.
Panel (b) shows the computational speedup, calculated as the ratio $\frac{M_{\rm FS}}{M_{\rm SD}}$ of the number of Brownian steps required to achieve a given uncertainty by each method.
We chose $M_{\rm FS} = \num{1000}$, thus the above expression conceptually reduces to how fast the SD method reaches the same accuracy as FS when \SI{100}{\percent} of the trajectory is used for the latter.

\indent We make the following observations for this homogeneous single-component model example.
(i) The off-diagonal elements of $\Gamma(\tau_n)$ that are set to zero are in fact distributed with a zero mean, confirming that they can be regarded as noise (Section 4 of the Supplemental Material\cite{molinari2020supplemental}).
This numerically confirms our theoretical proof that the SD method is unbiased.
(ii) Panels (a) and (b) show that the SD method produces more significant improvement in the weak-to-moderate correlation regimes, $0.5<f_c<1.5$. For strongly correlated single-component systems, the SD approach does not provide a considerable advantage over the FS method. Physically, in the latter regimes fluctuations of the ionic flux comprise the whole system, thus the total center of mass displacement of the system should be used to compute transport properties.
(iii) The advantage of the SD method over the FS grows for larger system sizes.
This is because the fluctuations of the center of mass decrease for larger systems, increasing estimate variance and thus requiring longer trajectories to converge.
(iv) Crucially, the SD method \textit{always} outperforms the FS method, as, for both heatmaps, the values never fall below \num{1}.
Rigorous justification of this point, and derivations of the proofs of variance reduction are given in Section 3 of the Supplemental Material\cite{molinari2020supplemental}.
The last observation implies that the SD approximation can be applied to any system without prior assumptions on the correlation structure.

{\subsubsection{\label{sec:level3}Lennard-Jones Liquid}
\indent We illustrate our approach by studying the physical meaning of the eigenvectors of $\langle\mathcal{C}_{ij}(\tau_1)\rangle$.
Additionally, Section 7 of the Supplemental Material reports the calculation of $D$ for a Lennard-Jones (LJ) interacting molecular system.
We perform molecular dynamics (MD) simulations to generate atomic trajectories of one dimer and one trimer immersed in a bath of \num{500} monomers. \\
\indent \autoref{fig:fig2_mainText} presents the study on the meaning of the eigenvectors of $\langle\mathcal{C}_{ij}(\tau_1)\rangle$.
\begin{figure}[h!]
    \centering
    \includegraphics[scale=1.0]{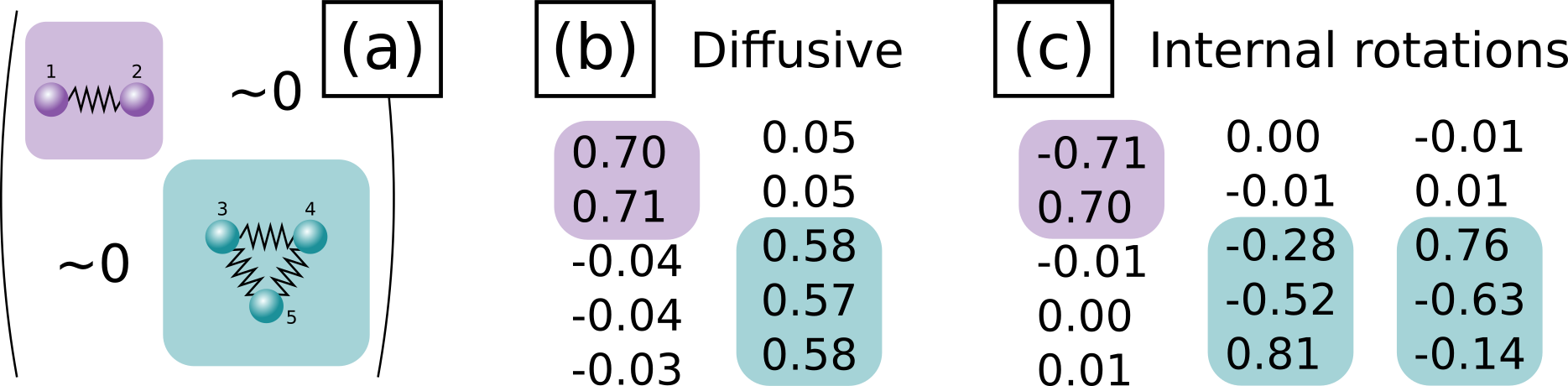}
    \caption{Lennard-Jones liquid diffusion results: (a) structure of $\langle\mathcal{C}_{ij}(\tau_1)\rangle$ for a single dimer and trimer in a bath of monomers; (b) and (c) covariance matrix eigenvectors.}
    \label{fig:fig2_mainText}
\end{figure}
Panel (a) represents the structure of $\langle\mathcal{C}_{ij}(\tau_1)\rangle$, consisting of three components: the diagonal, the inter-species off-diagonal, and the intra-species off-diagonal entries.
In this dilute case the inter-species off-diagonal entries of (a) converge to zero.
On the contrary, the bonded atoms constituting the dimer and trimer have strong immutable correlation as they are bound to move together.
As a result, $\langle\mathcal{C}_{ij}(\tau_1)\rangle$ has the structure of a block diagonal matrix, as sketched in~\autoref{fig:fig2_mainText}(a), and the SD approach reduces to the Nernst-Einstein description in the basis of molecules, which is equivalent to the cluster Nernst-Einstein method \cite{france2019correlations}.
Correspondingly, the eigenvectors of $\langle\mathcal{C}_{ij}(\tau_1)\rangle$ are partitioned into two sets.
One set corresponding to the diffusive drift of the center of mass of each molecule (b), and another set corresponding to non-diffusive intra-molecular motions, rotations and vibrations (c).
Specifically, for the dimer (highlighted in purple), there are two eigenvectors.
The first eigenvector, approaching $\left( \frac{1}{\sqrt{2}}, \frac{1}{\sqrt{2}}, 0, 0, 0 \right)$, describes the center-of-mass motion of the duplet, while the second, $\left( \frac{-1}{\sqrt{2}}, \frac{1}{\sqrt{2}}, 0, 0, 0 \right)$, corresponds to internal motion not contributing to diffusion and has vanishing eigenvalue in the limit of infinite statistics.
In general, any eigenvector whose eigenvalue and vector components sum approach zero does not contribute to diffusion.
Thus, eigenvectors of $\langle\mathcal{C}_{ij}(\tau_1)\rangle$ have intuitive physical interpretation as collective diffusion modes of the system, and form an efficient basis for analysing diffusive transport.
This method thus additionally provides an unsupervised automatic way to identify diffusing clusters and molecules in the case of strong short-range correlations, from only atomic motion without any prior information.}

{\subsubsection{\label{sec:level3}Electrolyte Conductivity}
As a realistic test of our method, we calculate the conductivity $\kappa$ for two battery electrolyte systems: 1. a lithium salt \li\tfo in an ionic liquid (IL), \emim\tfo, and 2. an amorphous lithium phosphate ceramic, \phosphate.
For the former, we also investigate the improvement in standard deviation of the estimate as a function of temperature.
Additionally, Section 8 of the Supplemental Material reports the same analysis for a highly-correlated ($f_c \approx 4.4$), garnet, \llzo.
{As expected from~\autoref{fig:fig1}, the SD applied to the highly-correlated garnet reduces to the FS for both computational cost and estimation.}
The above systems have been shown to exhibit significant ion-ion correlation both theoretically\cite{molinari2019general, marcolongo2017ionic, he2017origin} and experimentally\cite{gouverneur2018negative, kuwata2016lithium, rosenwinkel2019lithium}, with $f_c$ from 0.6 (for ionic liquids) to as high as 5 (for garnets).
\begin{figure*}[h!tb]
    \centering
    \includegraphics{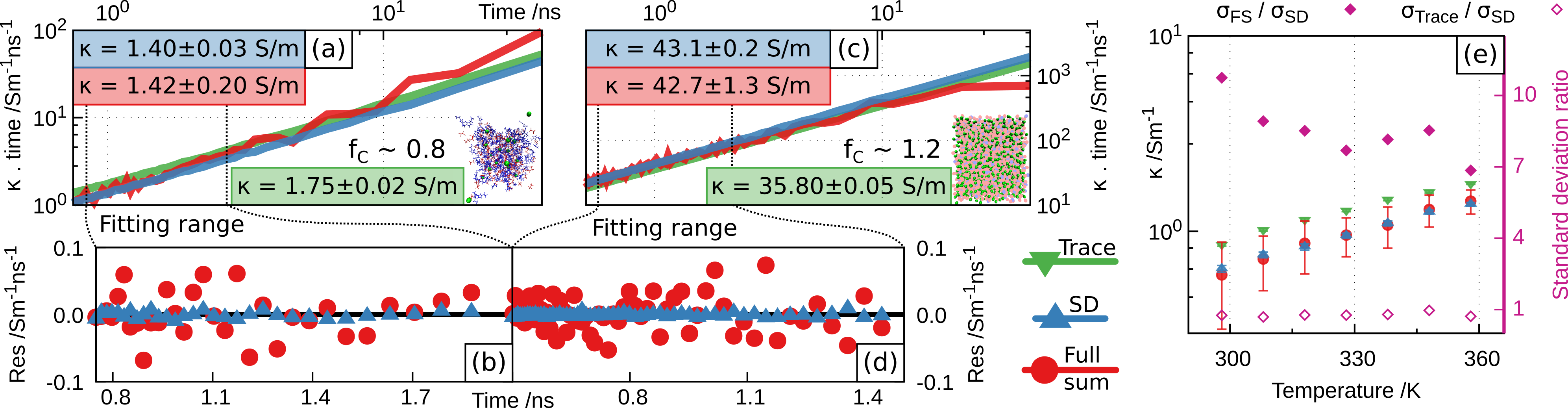}
    \caption{SD method applied to the calculation of $\kappa$ for two electrolyte systems: \li\tfo~in \emim\tfo~ionic liquid (a,b,e), and amorphous \phosphate~(c,d). (a) and (c) show $\sum_{ij} \langle\mathcal{C}_{ij}(\tau)\rangle$ as computed with the three different methods, (b) and (d) show the residuals over the time range used to compute $\kappa$, and (e) shows the Arrhenius plot for the ionic liquid system, as well as the SD variance reduction ratio.}
    \label{fig:realsystems}
\end{figure*}
The computed $f_c$ are \num{0.8} and \num{1.2} for the ionic liquid and solid state electrolyte, respectively. Consequently, the MSD method would result in a \SI{20}{\percent} overestimation and underestimation, respectively. \\
\indent We perform MD simulations and analyze the atomic displacement correlations to compute $\kappa$ with the three methods: trace, FS, and SD.
The ionic liquid-based electrolyte is composed of \num{178} 1-Ethyl-3-methylimidazolium (\emim), \num{19} \li, and \num{197} trifluoromethanesulfonate (\tfo) molecules, leading to a \li-salt molar fraction of \num{0.1}.
The interatomic potentials, the structure generation and equilibration protocols are inherited from our previous works\cite{molinari2019transport, molinari2019general, fadel2018effects}.
{For \phosphate, we create a supercell of the crystalline structure with \num{3456} atoms, \num{1296} of which are \li; the force-field is from \cite{pedone2006new}.
We note that, while the force-field from Pedone et al compromises between computational cost and accuracy, the SD method can be equally applied to position-position covariance matrices obtained with any energy model, provided a diffusive timescale can be reached.}
The simulated temperatures are \SI{358}{\kelvin} for the ionic liquid and \SI{600}{\kelvin} for \phosphate.
Full details in Section 6 of the Supplemental Material\cite{molinari2020supplemental}.
\indent Panels (a) and (c) show the drift over time of $\sum_{ij}\langle\mathcal{C}_{ij}\left(\tau\right)\rangle$ for the trace (green), FS (red), and SD (blue) approaches.
As in the other models discussed above, the displacement correlation $\sum_{ij}\langle\mathcal{C}_{ij}(\tau)\rangle$ is significantly less noisy for the SD method compared to the FS approach, and this translates to lower residuals, \autoref{fig:realsystems}(b,d) for both systems, with $\kappa$ matching the FS values.
The uncertainty of the $\kappa$ estimate is reduced by $\sim\SI{70}{\percent}$ for both electrolyte systems. \\
\indent Finally, \autoref{fig:realsystems}(e) shows the Arrhenius plot for the ionic liquid system, as well as the ratio in standard deviation between the SD and FS and trace, full and empty diamonds, respectively.
In this wide temperature range, while providing unbiased estimates of the fully-correlated conductivity, the SD approach outperforms the FS as it reduces the uncertainty on the estimate by an average of \SI{90}{\percent}, providing performances comparable to those of the Trace method as $\frac{\sigma_{\rm Trace}}{\sigma_{\rm SD}}\approx 1$.
A systematically lower standard deviation for all temperatures will also yield a better estimation of the activation energy since it is the slope of the Arrhenius plot.}

\section{\label{sec:level1}Conclusions}
In summary, we provide a superior approach capable of reducing the uncertainty of conductivity estimates of correlated systems. This is achieved by leveraging the correlation information encoded in the well-converged short-time position-position covariance matrix. 
The spectral analysis of the position-position covariance matrix is shown to be an unsupervised way to uncover stable collective diffusion modes and particle clusters, automatically revealing the microscopic physical mechanisms underpinning ionic transport in complex systems, without prior information as required for previously available methods.
Consequently, it enables accurate estimates of transport properties from significantly shorter molecular dynamics trajectories, by several orders of magnitude for larger systems, while capturing the full correlation contribution of the total flux, exact full summation approach.
We derive formal justification that the results are unbiased and provide rigorous bounds on the reduction of the variance of the estimates.
In addition, we numerically demonstrate the improvement and applicability of our approach on controlled models and two realistic electrolyte systems: \li\tfo~in \emim\tfo~ionic liquid-based and \phosphate~solid-state battery electrolytes.
These results open the possibility of rapid investigation of transport characteristics in complex concentrated electrolytes where correlation effects cannot be neglected.

\section{\label{sec:level1}Acknowledgments}
\indent We acknowledge useful discussions with Eric R. Fadel. N.M. is supported by the US Department of Defense MURI under Award No. N00014-20-1-2418. Y.X. is supported by the US Department of Energy (DOE) Office of Basic Energy Sciences under Award No. DE-SC0020128.
\bibliography{ProjectionConductivity.bib}
\end{document}